# Scheme for a spin-based quantum computer employing induction detection and imaging

Aharon Blank

*Schulich Faculty of Chemistry*

*Technion – Israel Institute of Technology*

*Haifa, 32000*

*Israel*

Phone: +972-4-8293679

Fax: +972-4-829-5948

Email: ab359@tx.technion.ac.il

URL: http://mr-lab.technion.ac.il/

Abstract

A theoretical spin-based scheme for performing a variety of quantum computations is presented. It makes use of an array of multiple identical "computer" vectors of phosphorus-doped silicon where the nuclei serve as logical qubits and the electrons as working qubits. The spins are addressed by a combination of electron spin resonance and nuclear magnetic resonance techniques operating at a field of ~3.3 T and cryogenic temperatures with an ultra-sensitive surface microresonator. Spin initialization is invoked by a combination of strong pre-polarization fields and laser pulses, which shortens the electrons' $T_1$. The set of universal quantum gates for this system includes an arbitrary rotation of single qubits and c-NOT operation in two qubits. The efficient parallel readout of all the spins in the system is performed by high sensitivity induction detection of the electron spin resonance signals with one-dimensional imaging. Details of the suggested scheme are provided, which show that it is scalable to a few hundreds of qubits.

## 1. Introduction

There are numerous suggested schemes for the implementation of quantum computers, ranging from calcium ions, superconducting qubits, color centers in diamonds, and a variety of nuclear- and electron-spin systems [1]. Each method has its pros and cons, but one of the leading candidates is based on the use of electron and nuclear spins in phosphorus-doped $^{28}$Si crystal ($^{28}$Si:P). The long coherence time of the electrons and nuclei in this system (in the range of 1 second [2]) compared to the short interaction and manipulation times of the electron spins (in the range of 10-1000 nanoseconds), puts it on a par with high vacuum ion traps and superconducting qubits. Furthermore, the ability to achieve real quantum entanglement with these systems makes them very attractive for quantum computation [3]. However, despite recent advances in P-doped purified $^{28}$Si fabrication methods [4], and the implementation of qubit operations [3], there remain some major obstacles in the experimental realization of a scalable spin-based quantum computer. The main problems are related to initializing the spins to a single quantum state, addressing and manipulating single spins out of the sample, and the greatest challenge of all – reading out the state of the individual spins after the calculation has ended.

Here we show that the use of conventional induction-detection electron spin resonance (ESR) spectroscopy and imaging techniques can lead to a solution for all of these problems and thereby enable the realization of a spin-based scalable quantum computer. Induction-detection (or Faraday detection) is the only method used in conventional magnetic resonance for spectroscopy and imaging experiments. It is very flexible, capable of "talking" to many individual spins in parallel, and also of detecting and imaging them efficiently as a whole using spatially-encoded individual information. In fact, it is only thanks to the induction method that true quantum entanglement has recently been demonstrated in a spin-based system [3]. All these, in principle, make induction-detection and imaging an excellent choice for the most demanding quantum computation (QC) applications; however, traditionally they lacked sufficient spin sensitivity and spatial resolution to be seriously considered as a viable option.

In recent years we have continuously pushed the spin sensitivity of induction-detection ESR, starting at the commercial state-of-the-art level of ~$10^9$ spins/√Hz (i.e., for one second of signal averaging time) and up to ~$10^6$ spins/√Hz in our recent work with $^{28}$Si:P samples, measured at cryogenic temperatures and a static field of ~0.5 T [5,6]. These improvements were achieved with the development of a novel ~20×65-μm-sized surface resonator; when combined with an imaging system, they deliver a resolution down to the ~400-nm range [7]. The basic theory on spin sensitivity by induction-detection ESR [8] tells us that continuing along this line towards even smaller resonators (inner size of ~1×10 μm), operating at higher static fields (3.3 T ~93 GHz) will improve spin sensitivity by at least 2-3 additional orders of magnitude [9]. The improvement in sensitivity ($10^3$-$10^4$ spins/√Hz) can be followed by a similar advance in spatial resolution (down to a few nanometers), as will be shown here. This will enable induction-detection ESR and imaging to surpass the sensitivity and resolution thresholds, making it a viable, important, and possibly a leading factor in the field of quantum computing. Achieving electron spin sensitivity and image resolution of this magnitude will enable the application of our new capabilities to a novel type of QC scheme, outlined below, which is specifically tailored for induction detection and could offer in the future a viable scalable approach for a truly useful quantum computer.

## 2. Scheme for $^{28}$Si:P-based QC with induction-detection ESR

We will now outline our suggested scheme for a $^{28}$Si:P–based QC, which employs both conventional pulsed ESR and nuclear magnetic resonance (NMR) spectroscopy tools along with induction principles for spin manipulation and readout. Our assumption is that a detection sensitivity of $\sim 10^3$-$10^4$ electron spins/√Hz is available for electron spins of $^{28}$Si:P at cryogenic temperatures ($\sim$4.2-10 K) and a static field of $\sim$3.3 T. As we shall show below, beyond a certain threshold the exact experimental sensitivity is not of paramount importance and will only affect the measurement ("calculation") time of the process. Our proposed QC scheme is outlined in Fig. 1 and its main details are described in the figure caption. It makes use of ideas originally suggested for N@C$_{60}$-based QC [10] combined with notions developed for Si:P-based QC [11], and some additional novel concepts that make use of our ultra high sensitivity induction-detection and high resolution imaging capabilities.

The general requirements for implementing a quantum information processing algorithm on a physical system are usually considered in terms of the DiVincenzo criteria [12], which include: (i) a scalable physical system with well-characterized qubits; (ii) the ability to initialize the state of the qubits to a simple fiducial state such as |000 . . .>; (iii) long relevant decoherence times, much longer than the gate operation time; (iv) a ''universal'' set of quantum gates; and (v) a qubit-specific measurement capability. We will now review these 5 requirements and show how the suggested scheme complies with them.

We start with requirement (i), which is met by a scalable system of qubits like the one shown in Fig. 1. Such an array of phosphorus atoms in silicon can be produced with Angstrom-scale precision by a combination of scanning probe lithography and high purity crystal growth [4]. Each phosphorus nucleus in the crystal serves as a logical quantum bit (qubit), while its adjacent electron is the working qubit. The array has two lattice constants: a short one (a) with n dopants that enables electron spins to interact through dipolar couplings along this linear vector; and a long one (b) with m dopants along the line that separates many identical copies of the same individual vector "computers". Typically m would be ~100 to enable having enough identical spins for signal averaging. This means that the array would extend to a distance of ~10 μm along the crystal's y-axis, which conforms to the larger dimension of a high sensitivity 93-GHz surface resonator with inner dimensions of 2×10 μm which we recently designed [9]. The arguments for the chosen spacing between the spins will be addressed below. The value of n, which is the number of qubits in a "computer", can be ~420, as dictated by both the smaller dimensions of the surface resonator (~2 μm) and the ~41.9-Gauss hyperfine separation between the two ESR signals for the two phosphorus nuclei states [13]. The latter issue is a limiting factor since the spins can be separated by linear gradients of 0.01-0.02 G/nm (1000-2000 T/m - see Appendix), where each spin is left with an individual "domain" to operate of ~0.1 G (i.e., 0.1G×420~42 G), which is sufficiently more than the homogenous linewidth of this electronic system [2]. Finally, it should be noted that the multiplicity of identical "computers" in our schemes can be used to eliminate the need for quantum error correction due to ransom spin flips, since the measured result averages over ~100 spins per qubit. (However, methods to correct for pulse imperfections, such as phase cycling, should still be employed.) Therefore, quantum error correction issues will not be considered here.

Criterion (ii), in its simplest form, requires just starting from a well-defined state [12]. This can be achieved here by placing all the phosphorus nuclei in their ground state. This can be realized by first initializing all the electron spins and then transferring their spin polarization to the nuclei using either dynamic nuclear polarization methods or algorithmic cooling from electron to nuclear spins, as was demonstrated recently in this exact system [3]. Ideally, the process of electron spin initialization should be achieved almost naturally due to the high static magnetic field and low temperature of operation. However, the ratio between the populations of the electron spin energy levels is not low enough under our proposed experimental conditions ($\frac{N_u^e}{N_d^e} \equiv \alpha = e^{-\hbar\omega_0/k_B T}$ ~0.34 at a field of 3.3 T and 4.2 K). We can make use of our suggested scheme to overcome this difficulty in the following manner: a DC pulse of ~25 A that flows in the resonator's copper loop structure generates a static field of ~6 T in its center; aAnother 6 T can be added by using the microwires on the sample itself with an additional current drive of ~25 A (see Appendix), resulting in a total pre-polarizing field of ~12 T that reduces $\alpha$ down to ~0.02. The duration of such pre-polarization pulse can be minimized by the use of a laser light pulse that was found long ago to reduce the electrons' $T_1$ by almost 3 orders of magnitude at 1.2 K [14] and can probably reduce it by an even larger factor at higher temperatures. Here we assume that under laser irradiation the electrons' $T_1$, and therefore the pre-polarizing time, are both typically ~10 μs at our device's relevant temperatures of operation (4.2-10 K). The high copper conductivity at cryogenic temperatures (>1.5×10$^{10}$ S/m) and the high silicon and copper thermal conductivity minimize the resonator's heat dissipation during this powerful current pulse and facilitates the use of such a trick (see Appendix).

Criterion (ii) in its more advanced form requires a continuous, fresh supply of qubits in a low entropy state (like the |0> state), but these are primarily needed for quantum error correction schemes, which, as noted above, are not considered

in our approach. In any event, as noted also below, some of our quantum calculations schemes require spin initialization that can be done in a relatively short time using the laser-assisted scheme outlined above.

The third DiVincenzo criterion is addressed through the unique long coherence times of the $^{28}$Si:P system, which, as noted above, approach 1 second at ~3 K for electrons and are even longer for nuclei [2]. (These coherence times refer to bulk samples with small concentrations down to $10^{14}$ electron spins in 1 cm$^3$.) However, the coherence time alone is not enough and what really is important is the ratio between the coherence time and the "clock time" of the quantum computer (the execution time of an individual quantum gate), which must be in the order of $10^4$-$10^5$ [12]. This ratio will be estimated for the proposed unique sample topology by estimating the relevant coherence times and then taking into consideration the typical time to execute a quantum gate.

A two-dimensional array of phosphorus atoms, such as the one shown in Fig. 1, was not fabricated and its coherent times were certainly not measured. In order to estimate the coherence time of such an array under the possible influence of large static field gradients, we consider all relevant relaxation mechanisms ranging from small to large, as explained below:

(a) Spin-lattice relaxation: An obvious upper limit to the coherence time is due to spin-lattice relaxation mechanism. It is known from early papers that this relaxation rate has a steep temperature dependence that can be explained via spin-phonon interaction processes [14,15]. As a result, $T_1$ has a very week dependence on donor concentration and on the isotopic content of the silicon [16]. At the temperature range of 4-10 K, which is the relevant regime in our proposed scheme, $T_1$ changes from ~ 1 ms to ~ 10 s and sets the upper limit for the coherence time [16,2].

(b) Spectral diffusion due to dipolar fluctuations of $^{29}$Si nuclear spins: This interaction gives rise to a temporally random effective magnetic field at the localized electron spin, leading to irreversible decoherence. Such mechanism has been shown to have a significant effect on the electrons' coherence time, ranging from microseconds for a 100% $^{29}$Si sample up to a few milliseconds in a 0.08% $^{29}$Si sample, for ~ $10^{15}$ P dopants in 1 cm$^3$[17]. For this interaction, $1/T_m$ can be estimated to be equal to the interaction energy between the electron and the nuclear spin, $D=15.7/r^3$, where $r$ is in nanometers and $D$ in kilohertz (using the point-dipole approximation for the interaction energy). Indeed, at the lower regime of the $^{29}$Si concentration, the typical distance between the $^{29}$Si nucleus and the electron spins is ~2.5 nm, resulting in an estimated $T_2$ of ~1 ms, which a bit underestimates the experimental value of a few milliseconds found. Another approximate expression [18]: $\frac{1}{T_m} \approx 0.37 \gamma_e^{1/2} \gamma_{^{29}si}^{3/2} Nh [0.5 \times 1.5]^{1/4}$, gives $T_m$~85 ms for 0.08% $^{29}$Si sample, which overestimates the experimental results by a factor of ~10. In any case, all this means that further reduction of the $^{29}$Si concentration does not provide a significant improvement in the coherence time, when the defect concentration is ~$10^{15}$ spins/cm$^3$, since then the electron spins' interaction becomes the dominant mechanism (see next item) [2].

(c) Electron spins spectral diffusion: This decoherence mechanism arises from electron spin flip-flops of nearby donor pairs [2] (so-called indirect flip-flop processes [19]). The contribution to decoherence due to this mechanism, $1/T_m$, is roughly equal to the interaction energy between two electron spins, $D=12.98 \times 4/r^3$, where $r$ is in nanometers and $D$ in megahertz. Thus, along the large dimension of our array, the flip-flop process limits the relaxation time to ~20 ms for an interspin distance of ~100 nm. (This estimation is in agreement with the experimental results presented above that found, at low temperatures, $T_2$ of a few milliseconds for a P concentration of $10^{15}$ spins/cm$^3$, corresponding to an average interspin distance of 100 nm.) However, along the condensed dimension, $1/T_2$ can reach values in the range of 50-400 kHz, which is essentially the interaction rate. Therefore, there seems to be a contradiction between the need to increase the coherence time and at the same time to reduce the interaction time between the spins. Luckily, this can be resolved by applying a field gradient along the condensed dimension of the array that effectively eliminates this flip-flop mechanism [2]. A gradient of 0.01-0.02 G/nm is sufficient for the elimination of such relaxation, and it can be easily obtained from two microwires separated by ~2 μm (Fig. 1) with ~5.5-11 mA of current flow in them (see Appendix).

Other potential decoherence mechanisms can involve dipolar interaction with surface or buried paramagnetic defects (e.g., broken bonds or point defects). However, these can be largely eliminated by placing the phosphorus atoms at a depth of more than ~100nm from the surface and using a high purity silicon with low numbers of point defects in its volume (less than ~$10^{14}$/cm$^3$ is readily available).

We can therefore summarize this part and say that for an interspin distance of ~5 nm (interaction rate of ~400 kHz), the typical quantum gate operation would take a few microseconds (with the pulse sequences described below), which is good but a bit short of the DiVincenzo criterion considering the electrons' estimated $T_2 \sim 20$ ms (due to the spin-spin interaction along the sparse dimension of the array). However, since the coherence of the electrons can be transferred to their adjacent phosphorus nuclei (see below), which have a $T_2$ of more than one second [11], it is possible to have almost $10^6$ quantum gate operations during one single coherent process of our nuclei qubits.

One of the most challenging aspects is the fourth (iv) criterion, which requires having a "universal" set of quantum gates. In principle, this means that the suggested scheme should support quantum algorithms that are essentially a set of unitary transformations, $U_1$, $U_2$, $U_3$…, each acting on a small number of qubits (typically not more than 3). These unitary transformations constitute the required quantum calculation. The implementation of these transformations in the physical world can be done by identifying Hamiltonians which generate these unitary transformations $U_1 = e^{iH_1 t/\hbar}$, $U_2 = e^{iH_2 t/\hbar}$ …, and then, ideally, our scheme should be able to turn on and off these Hamiltonians in a serial manner during the calculation period. In practice, however, there is no need to use a great variety of transformations to implement a general quantum computation. A universal set of transformations can be used that includes only the ability to perform arbitrary rotations of single qubits and the ability to implement a controlled-NOT (c-NOT) gate for two qubits.

At this point we must go into a detailed quantum description of the system to show how these operations can be carried out. Let us start by considering single-qubit transformations using our proposed scheme. Here we closely follow the work described in reference [20] that deals with a similar case. Since we apply a strong field gradient of ~0.01-0.02 G/nm we can address individual electrons along the chain and, at least initially, consider each phosphorus atom and its unpaired electron individually. (Nevertheless, further into our treatment we will also periodically consider the effect of neighboring phosphorus atoms in a quantitative manner, using approximations). Thus, the relevant Hamiltonian of the problem is (using high field approximation):

$$H = g_e \mu_B B_0 S_z - \gamma_n B_0 I_0 I_z + A S_z I_z \qquad [1]$$

where $\mu_B$ is the Bohr magneton, $g_e$ the electron g factor, and $\gamma_n$ is the gyromagnetic ratio of the nucleus. $S_z$ and $I_z$ are the z components of electron and nuclear spin operators, **S** and **I**, respectively. The energy levels and the schematic NMR and ESR spectra of this system are shown in Fig. 2. For operation at a static field of ~3.3 T the ESR spectrum has two lines at ~93 GHz, separated by A~117.4 MHz, while the NMR signal has two transitions, at ~1.82 MHz and 115.5 MHz. As noted above, we use the nuclear spin to store the quantum state of the qubit. Therefore, every single-qubit transformation must start by first swapping the information from a specific nucleolus of relevance to its neighboring "working" electron spin. In principle, this can be achieved by one of the pulse sequences described in reference [21] in the context of NMR. However, close inspection of these sequences reveals that they require typical interpulse intervals of $\frac{\tau}{2} = \frac{\pi}{2A} = 13.5\ ns$, which are achievable for 93-GHz ESR but almost certainly impossible for radiofrequency (RF) pulses in the 115-MHz range. Luckily there is also an alternative approach that makes use of three consecutive π pulses selectively for the relevant transitions, as shown in Fig. 3 (up to a constant phase) [20,11]. This so-called "swap" operation from the nuclei to the electron can be carried out non-selectively, without any field gradient, on all the n spins along the computer chain (and also for all the m spins along the averaging axis). This means that the three π pulses can be completed in less than 1 μs: short ~10-ns pulses for 93-GHz ESR could be easily obtained with our miniature surface resonators even with <mW of power [9,6,8]. For the NMR pulse one can make use of a ~1-mm RF coil surrounding the sample that can generate a π pulse shorter than 1 μs with ~10 W of RF power. Following that nonselective swap operation, the gradient is turned on and a single electron along the computer chain (in parallel to its m-equivalent electrons along the averaging dimension) can be accessed via selective excitation microwave (MW) pulses to perform any required one-qubit rotation. These rotations would typically last no more than 100 ns, meaning that the effect of dipolar coupling from neighboring spins (operating at a maximum rate of ~400 kHz for a nanometer spin separation) would be negligible during this time frame. Since the state of the phosphorus nuclei near the electron is unknown, the same rotations should be repeated with a 41.9-G field offset (or MW frequency offset). A second nonselective swap operation without field gradient would bring us right where we started with the exception of the specific nuclei being rotated by the required one-qubit operation.

We now turn to two-qubit operations, which are more costly time-wise, but still short enough. Here we start with a specific state of the logic qubits (nuclei) at some specific stage during the requested quantum calculation. The two-qubit operation is then started by resetting all the electron spins with a laser light pulse and a strong field pre-polarization pulse (see Appendix), similar to the initialization stage of item (ii) in DiVincenzo's list. As noted above, this light-assisted reset process should transfer almost all the electron spins to the ground state in a time scale of ~10 μs. Following this we implement a selective swap operation that acts only on the pair of neighboring electron spins that will be used for the two-qubit operation, using a sequence similar to the one shown in Fig. 3. Selectivity with respect to the electrons is easy, but in order to be selective with respect to the nuclei a gradient of ~$10^6$ T/m has to be produced during the RF pulse; this is not easy but still doable with the microwires integrated into the sample (see Appendix). Such a gradient will make it possible, with an NMR pulse of ~5-10 μs, to selectively address only the two nuclei intended for the two-qubit gate operation and to swap their quantum state with their nearby electron spins. At the end of this stage we should have two electron spins that are situated in the required initial quantum state (identical to their neighboring nuclei), while all other electrons around them are in their ground state. We can write the Hamiltonian for these two electrons (denoted A and B) and their adjacent nuclei as:

$$H = g_e\mu_B B_0^A S_z^A + g_e\mu_B B_0^B S_z^B - \gamma_n B_0 I_z^A - \gamma_n B_0 I_z^B + A\left(S_z^A I_z^A + S_z^B I_z^B\right) + 2\pi D S_z^A S_z^B \quad [2]$$

where we took into consideration the different magnetic fields in site A and site B only for the electrons (it is negligible for the nuclei at ~1000 T/m). This Hamiltonian can be used to selectively implement a CNOT gate under our nominal gradient of 1000-2000 T/m (0.01-0.02 G/nm) with the two electrons using the sequence appearing in Fig. 4 [22]. It should be noted that this sequence effectively refocuses the hyperfine interactions from the electrons' neighboring nuclei and also the effects of neighboring electrons. (In any event, the neighboring electrons are all in their ground state so they merely shift slightly the resonance frequency of the "working" electrons but do not affect the outcome of the c-NOT operation). For an electron dipole-dipole interaction of $D$=100 kHz, the duration of this c-NOT gate would be ~5 μs. Following the completion of the c-NOT sequence, a selective swap operation is again carried out only on these two electrons and their adjacent nuclei, as described above.

The last step in the quantum calculation, also the fifth DiVincenzo criterion, is the readout of the state of all the spins. Traditionally, this has been the most demanding task in spin-based QC schemes. Induction detection, however, makes it rather straightforward. The states of the logical qubits are first swapped nonselectively with all the electron spins and then a Carr Purcell Meiboom Gill (CMPG) MW echo sequence is used under a static gradient of

~1000 T/m to collect the signal from all the electron spins in parallel [6]. A phase-corrected Fourier transform of the acquired signal would result in a one-dimensional real vector where positive amplitudes correspond to a certain spin state and negative ones to the opposite state. Each spin along the computer chain appears as a point in this one-dimensional image, and a strong-enough signal is obtained for individual spins since the signal is the sum of all the parallel identical m spins. A reasonable approach for spin readout is to commit approximately half of the available coherence time to signal averaging, meaning that over a typical 2 seconds of coherent "calculation" time (the nuclei $T_2$), one second is devoted to various logic gate operations and the other to collecting the output signal. This means that, with a sensitivity of ~1000 spins /√Hz (SNR=1), our proposed scheme will have to repeat every "calculation" for several minutes to obtain a reasonable SNR for the spin readout. There is no well-defined threshold for the required minimal spin sensitivity to enable such detection scheme, but clearly repeating the same calculation for a few hours will constitute a reasonable limit, meaning that spin sensitivity cannot be worse than ~$10^4$ spins /√Hz.

3.  **Summary and conclusions**

We have presented a new scheme for the possible physical implementation of quantum computation algorithms on a scalable spin-based system. The scheme makes use of several recent advances in the field of atomic-scale lithography, sensitive induction detection and imaging, and the generation of short powerful current pulses. The sample of $^{28}$Si:P has many unique features, such as long coherence times for the nuclei and the electrons, light-dependant $T_1$, a combination of electron and nuclear spins, and huge thermal conductivity, all of which contribute to the possibility of implementing the suggested scheme.

4. Acknowledgments

This work was partially supported by grant #201665 from the European Research Council (ERC).

5. **Appendix**: Calculation of magnetic field gradients, pre-polarization field and heat dissipation in the sample

In this appendix we provide more details about the method of generating the required magnetic field gradients and large pre-polarizing field, while taking into consideration the corresponding heat load generated in the sample. The magnetic field gradient is generated in our configuration by two parallel wires with a square cross section of a×a whose center points are separated by Δ, and it is given by the expression:

$$G(x) \approx \frac{\mu_0}{2\pi}\left[\left(\frac{1}{x+\Delta/2+a/2}\right)^2 + \left(\frac{1}{x-\Delta/2-a/2}\right)^2\right] \quad [T/m \cdot A] \qquad [3]$$

For nominal values of a=1 μm and Δ= 2 μm, this gives ~1.8×10$^5$ T/m for 1 Ampere of current. The capability of the wires to support the current depends on the latter's magnitude, duration, the power dissipated in the microwires, and the heat conduction/dissipation of the assembly. At low temperatures the electric conductivity of pure copper is at least 1.5×10$^{10}$ S/m [23], which means that the resistance of each wire section (that is at least 10-μm long) is R~0.6 mΩ. This implies that for a continuous 1 A of current the power dissipation in each wire section is ~0.6 mW (other wires leading to it can be much thicker and with negligible dissipation). -The thermal conductivity of copper at 8 K is ~2,500 [W/(m×K)] while the thermal conductivity of isotopically-enriched $^{28}$Si is

~10,000 [W/(m×K)] at the same temperature range [24]. This makes for very efficient heat dissipation, which boils down to a relatively minor increase in temperature when generating the field gradients. In order to verify this point we carried out a finite element simulation (using COMSOL Multiphysics 4.2a) that takes into consideration a helium vapor flow rate of 75 liters (of gas) per hour at 8 K (Fig. 5) in a typical cryostat tube with an i.d. of 50 mm. The simulation shows a steady state temperature increase of less than 0.5 K for a continued current of 1 A in each microwire. This means that for a gradient of 1,000 T/m that will be operated most of the time, the temperature increase would be much smaller (since it requires just a few milliamperes of current). Furthermore, for the gradient of $10^6$ T/m that is required in some of the operations (to separate the nuclei frequencies), the current is ~5.5 A, meaning that the power is ~30 times larger than what we calculated. However, such a high gradient will be applied only for a very short time and in a low duty cycle (< 1%), which means that it will be still well within the system's reasonable power-handling range, without significant changes in temperature.

In addition to the magnetic field gradient, we can make use of the surface resonator itself to generate a powerful pre-polarization static magnetic field as required by our QC scheme. This can be done by driving a strong current through it, as schematically presented in Fig. 6. Furthermore, one can use the same microcoils used for the field gradient but reversing the direction of the current in one of them. Due to current flowing in the microcoils or the surface resonator, the field at the sample position has similar dependence and similar contribution for a given DC current. The magnetic field for 1 Ampere of driving current in each microwire is given by the expression:

$$B(x) = \frac{\mu_0}{2\pi} \left[ \frac{1}{x + \Delta/2 + a/2} - \frac{1}{x - \Delta/2 - a/2} \right] \quad [T/m \cdot A] \qquad [4]$$

which gives a field of ~0.27 T for the typical wire parameters $a$=1 μm and $\Delta$=2 μm. This means that a drive of ~25 A in the microwires together with an additional 25-A drive into the surface resonator will generate a field larger than 12 T. Obviously, this current cannot be sustained for long periods of time. A typical pulse would be ~10-μs long, assuming that a synchronized laser pulse would reduce the electrons' $T_1$ to that time scale at ~8 K. We have recently developed an efficient scheme for generating such powerful short currents in small coils [25]. If the use of such pulses is kept to a minimum (i.e., one every ~10 ms on average, to initialize the spin state of the electron or for specific gate operations), then the average power deposited in the wires and resonator structure is ~1 mW, which is still within the acceptable values – see Fig. 5. Due to the extreme high thermal conductivity of copper and $^{28}$Si at low temperatures these figures can be improved if a larger sample slab (an effective heat sink) is considered.

**6. Figure captions**

Figure 1: The suggested QC scheme to be used in conjunction with ultra-high sensitivity/high-resolution induction detection (schematic view, not to scale). A two-dimensional array of phosphorus atoms is produced inside a pure $^{28}$Si single crystal, based on existing scanning probe lithography fabrication technologies [4]. The crystal is placed upside down on the center of our ultra-sensitive surface resonator [6,9] and operated at cryogenic temperatures. The insert in the upper left corner shows a picture of our present 20×65-μm (inner size) 17-GHz resonator. Each phosphorus nucleus in the crystal serves as a logical quantum bit (qubit), while its adjacent electron is the working qubit. The array has two lattice constants: a short one (a) that enables electron spins to interact through dipolar couplings along this linear vector (similar to the manner described in reference [10]), and a long one (b) that separates many identical copies of the same individual vector "computers". Individual spins can be addressed by applying a large magnetic field gradient with a DC current into microwires (separating the spins in the frequency domain) and the state of all the spins can be read out in parallel through a one-dimensional image along the x-axis of the crystal. All parallel identical "computer" vectors should give the same spin state, thereby increasing the measured signal and also eliminating the need for quantum error correction. Swapping information between working electron spins to logical nuclear spins can be carried out by combined radiofrequency (RF) and microwave (MW) pulse sequences, as described in reference [11].

Figure 2: The energy levels and the corresponding ESR and NMR transitions of an isolated $^{28}$Si:P system. ESR levels are marked in blue and NMR in red.

Figure 3: Schematic description of the "swap" operation of exchange of the quantum state between the nuclei and its nearby electron. The sequence includes three consecutive π pulses (up to a constant phase), operating on the electron (blue pulse), nuclear (red pulse), and then again electron transitions.

Figure 4: A MW pulse sequence for c-NOT operation with two electron spins. The thin pulses are π/2 pulses and the thick ones are π pulses with the phase of the pulses listed below them. Here, $\tau = \frac{1}{2D}$ (D in Hz).

Figure 5: Finite-element calculation of the velocity profile (a) and the temperature during a steady state operation of the microwires with 1 A of current flow starting from base temperature of 8 K for the helium vapor. The sample is described as a $^{28}$Si piece with dimension of 2×2×0.5 mm.

Figure 6: Schematic representation of our surface resonator with the possibility of running a DC current through it, enabling an efficient pre-polarization of the sample.

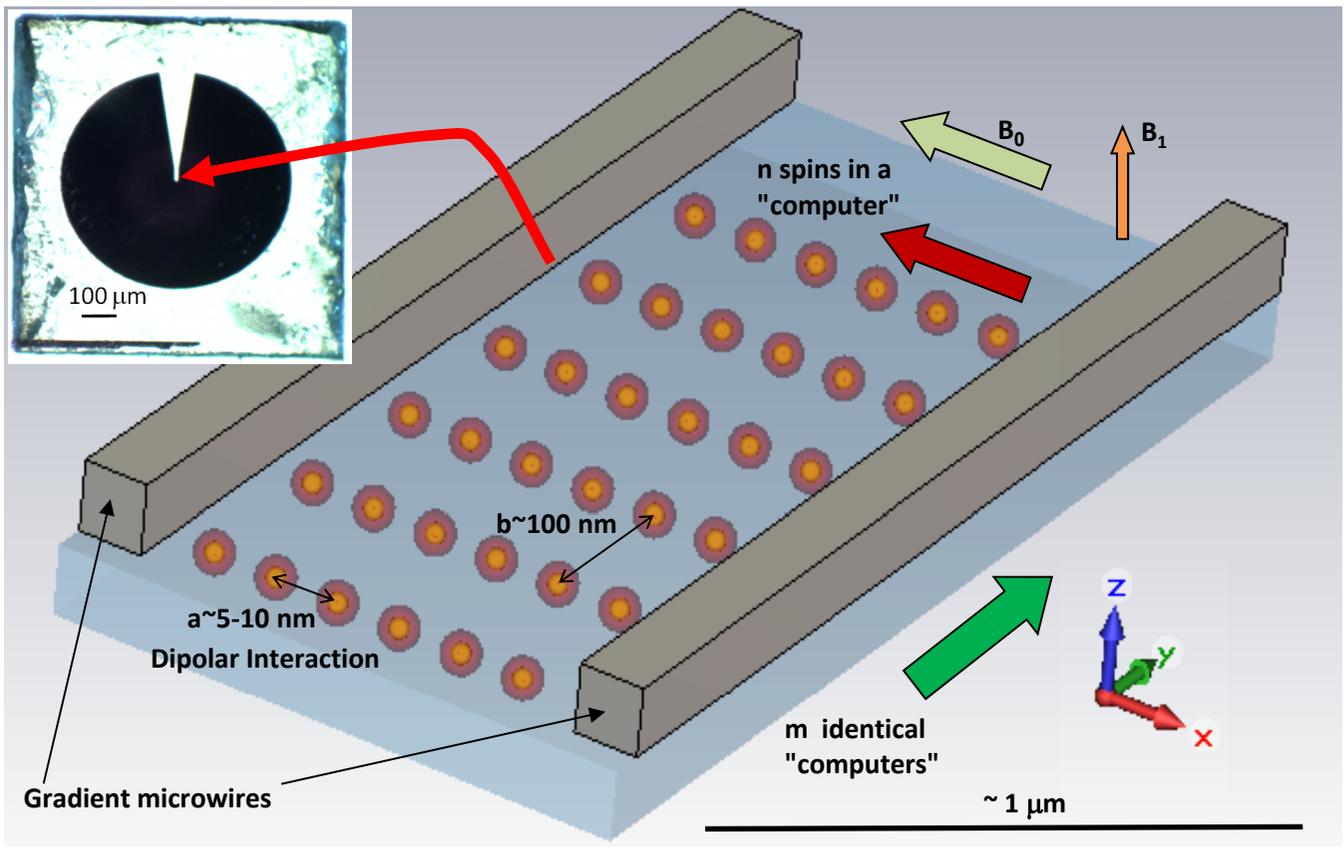

Figure 1

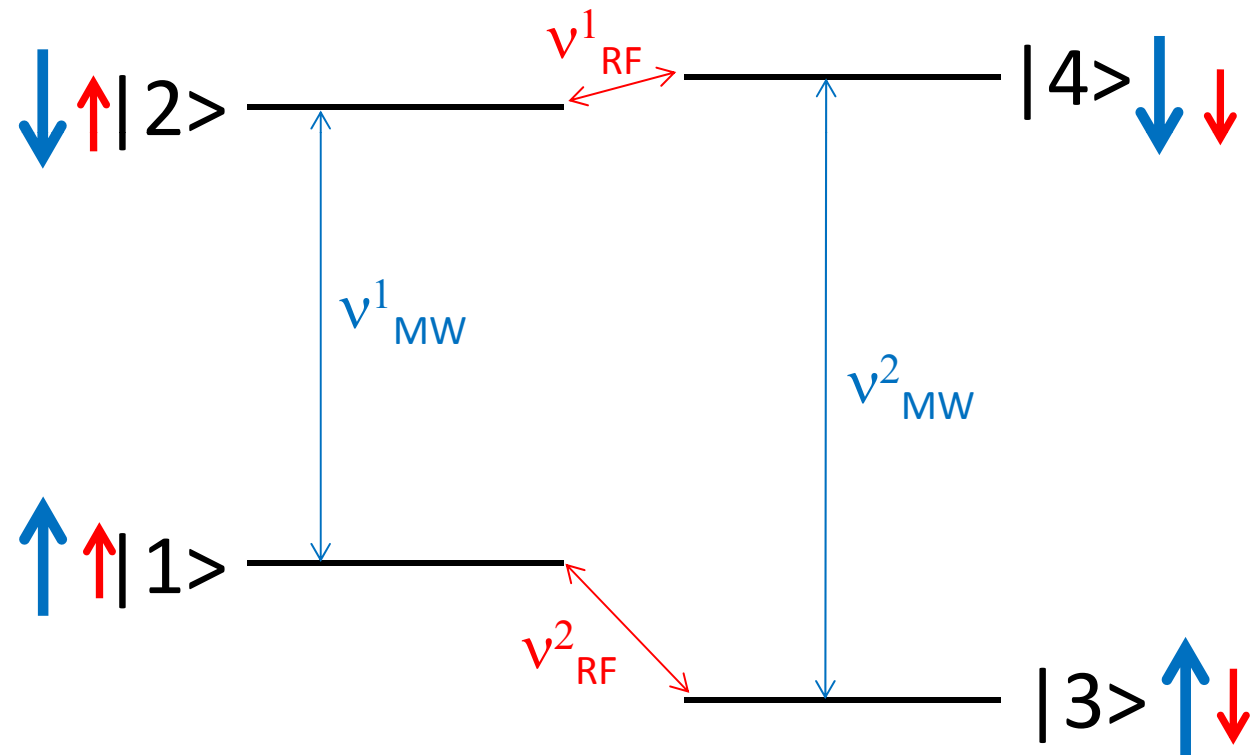

Figure 2

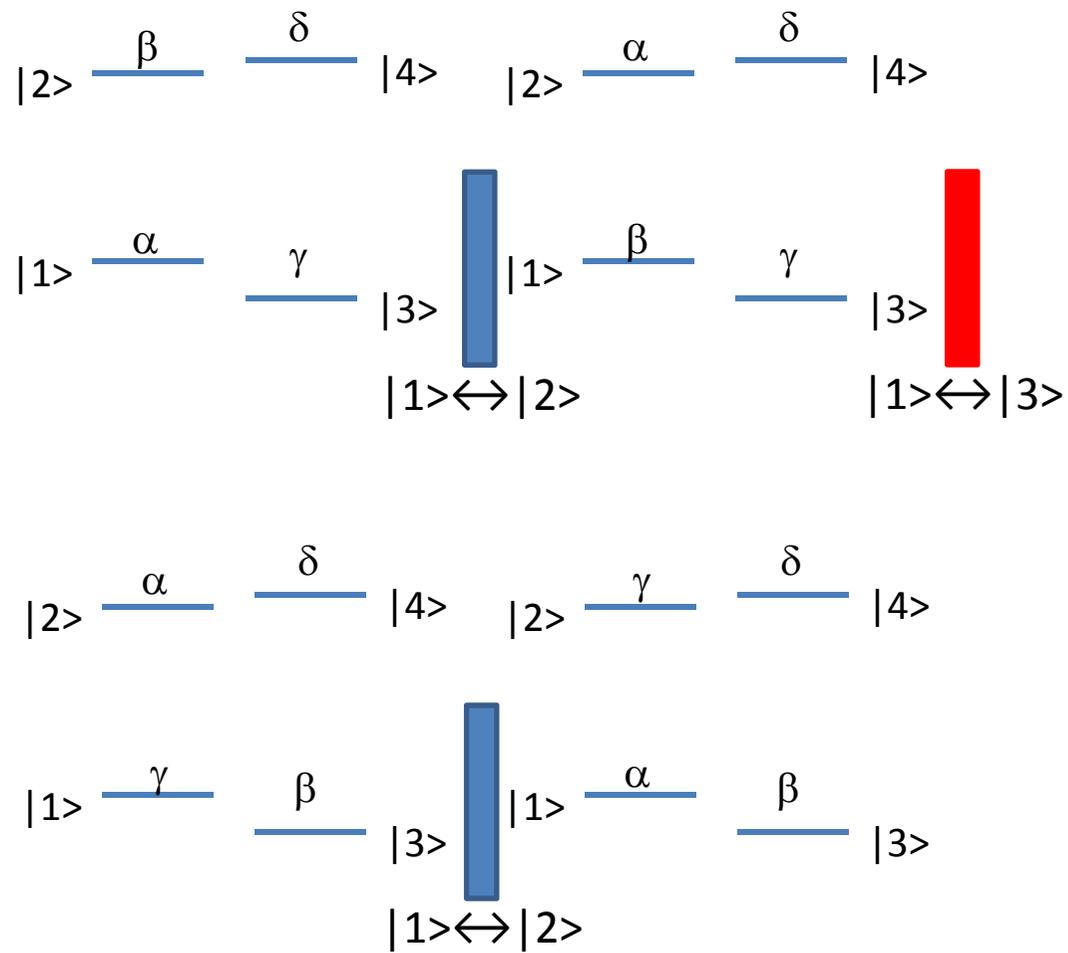

Figure 3

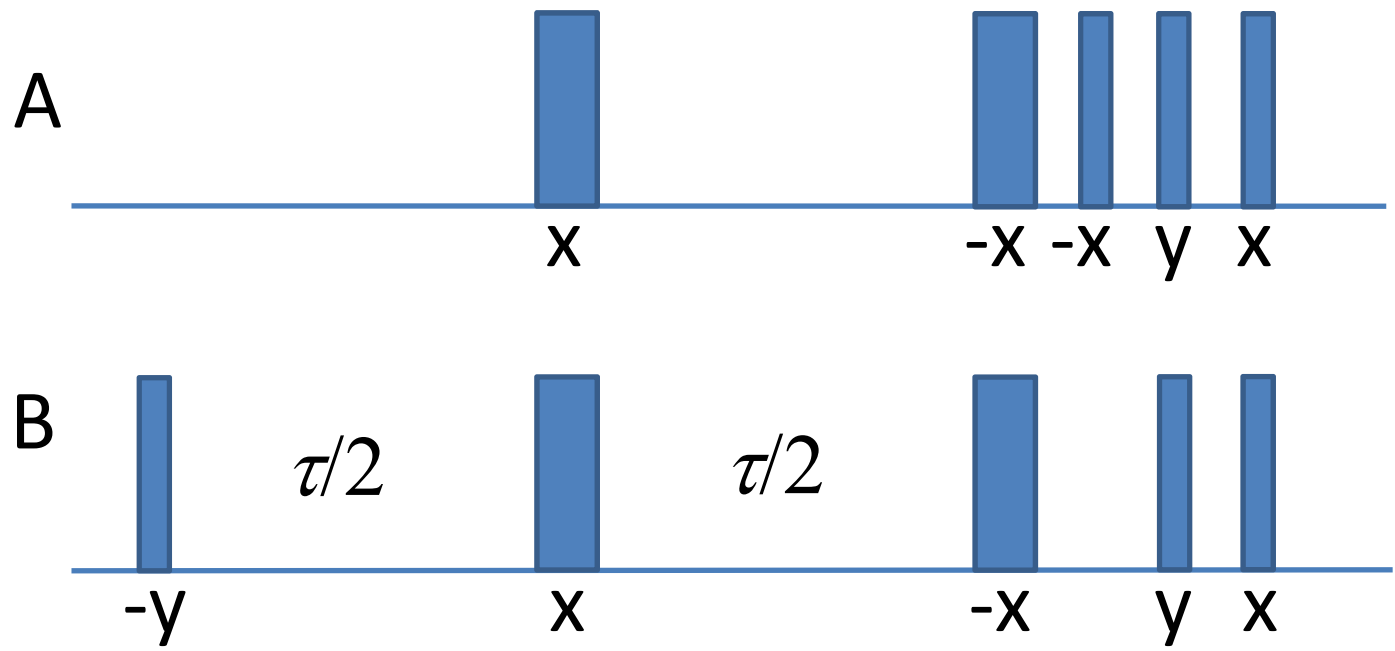

Figure 4

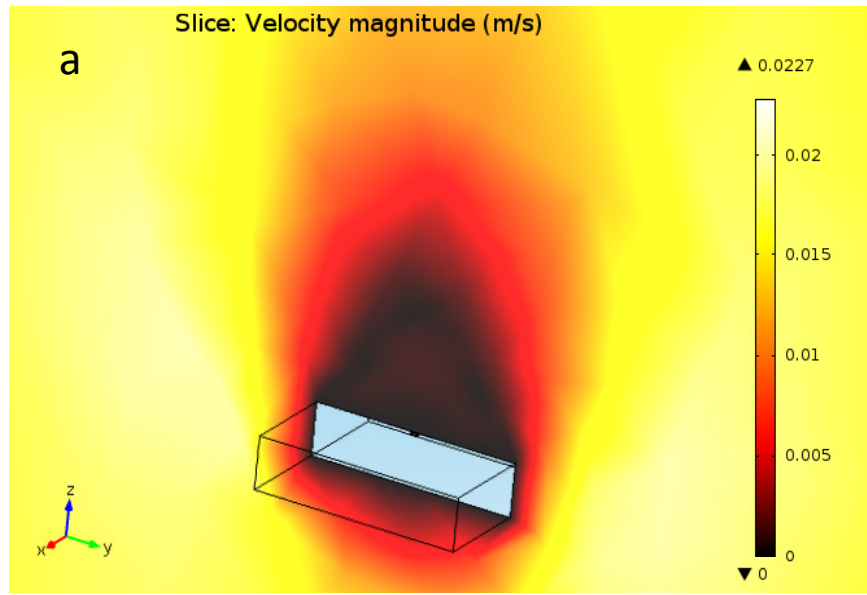
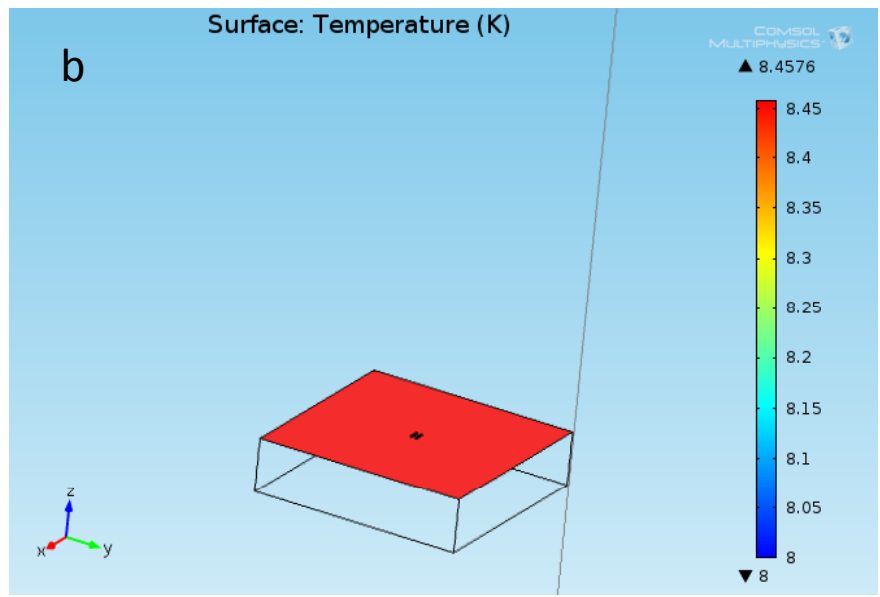

Figure 5

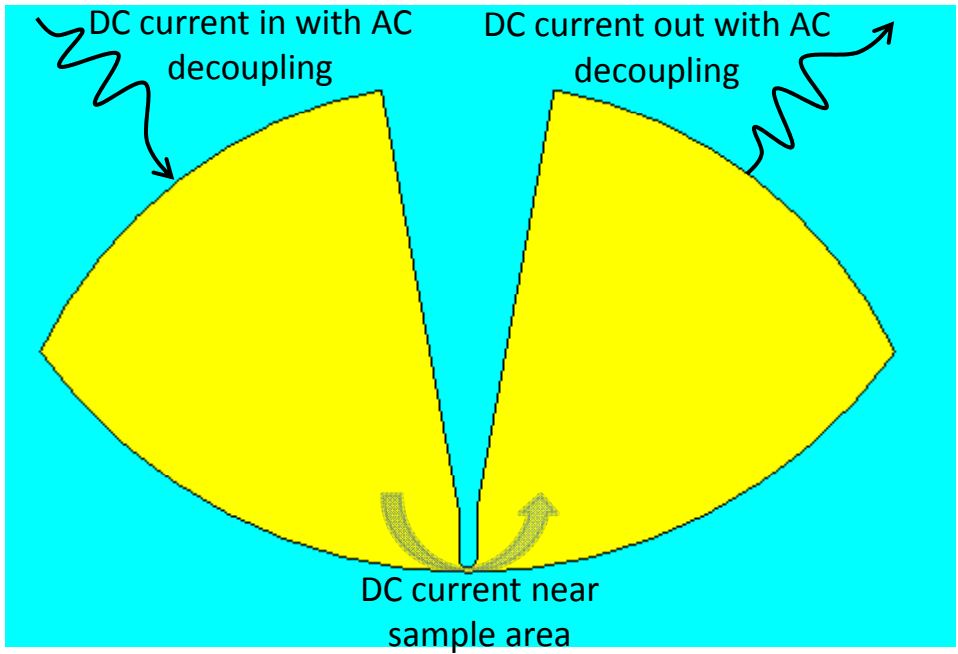

Figure 6